\documentclass[preprint,notitlepage,superscriptaddress]{revtex4-2}

\usepackage[utf8]{inputenc}
\usepackage{amsmath,amsfonts,amssymb}
\usepackage{graphicx}
\usepackage[toc,page]{appendix}

\usepackage{hyperref}
\usepackage[usenames,dvipsnames]{xcolor}

\usepackage{tikz}
\hypersetup{colorlinks=true, linkcolor=BrickRed, urlcolor=blue!50!black, citecolor=blue!50!black}

\usepackage[normalem]{ulem}

\begin{document}

\title{Optimal navigation strategy of active Brownian particles in target-search problems}

\author{Luigi Zanovello}
 \affiliation{Institut f\"ur Theoretische Physik, Universit\"at Innsbruck, Technikerstra{\ss}e 21A, A-6020, Innsbruck, Austria}
 \affiliation{Dipartimento di Fisica, Universit{\`a} degli studi di Trento, Via Sommarive 14, 38123 Trento, Italy}
\author{Pietro Faccioli}%
 \affiliation{Dipartimento di Fisica, Universit{\`a} degli studi di Trento, Via Sommarive 14, 38123 Trento, Italy}%
 \affiliation{Istituto Nazionale di Fisica Nucleare - Trento Institute for Fundamental Physics and Applications, Via Sommarive 14, 38123 Trento, Italy}
\author{Thomas Franosch}
 \affiliation{Institut f\"ur Theoretische Physik, Universit\"at Innsbruck, Technikerstra{\ss}e 21A, A-6020, Innsbruck, Austria}
\author{Michele Caraglio}
 \affiliation{Institut f\"ur Theoretische Physik, Universit\"at Innsbruck, Technikerstra{\ss}e 21A, A-6020, Innsbruck, Austria}
 \email[Electronic mail: ]{michele.caraglio@uibk.ac.at}

\date{\today}

\begin{abstract}
We investigate exploration patterns of a microswimmer, modeled as an active Brownian particle, searching for a target region located in a well of an energy landscape and separated from the initial position of the particle by high barriers.
We find that the microswimmer can enhance its success rate in finding the target by tuning its activity and its persistence in response to features of the environment.
The target-search patterns of active Brownian particles are counterintuitive and display characteristics robust to changes of the energy landscape.
On the contrary, the transition rates and transition-path times are sensitive to the details of the specific energy landscape.
In striking contrast to the passive case, the presence of additional local minima does not significantly slow down the active target-search dynamics.
\end{abstract}

\maketitle

\section*{Introduction}

Active matter has received increasing attention in the past two decades~\cite{roma2012,Lauga_2009,bech2016} and has evolved to an interdisciplinary field, involving physics~\cite{roma2012,Fodor2016} biology~\cite{Cates_2012,Needleman2017,Lipowsky2005}, medicine~\cite{Henkes2020,Fodor2018}, and robotics ~\cite{marc2013,Erkoc2019}.
Examples of active matter systems include natural microswimmers, \textit{e.g.} bacteria, sperm cells, or amoebae~\cite{lauga2006,eise2006,wu2016}, but also artificially synthesized agents, such as the renowned Janus particles~\cite{howse2007} or micro- and nano-robots~\cite{bech2016}. 
Active particles have become a paradigmatic model to study the non-equilibrium dynamics of these systems, such as collective motion~\cite{vicsek2012,elge2015} or motility-induced phase separation~\cite{cates2015}, but are also of interest for the plethora of potential applications that they present already at the single-particle level.

Exploration and target search in complex environments are central problems in many realistic scenarios for such active agents.
For living organisms these tasks are a matter of survival, for example, in bacteria foraging nourishment~\cite{bech2016} or for sperm cells searching a fertile egg~\cite{eise2006}. 
Artificial swimmers have been envisaged to perform specific purposes on the microscale, such as targeted drug delivery~\cite{naah2013,patr2013} or water decontamination~\cite{Vilela2017}.
In all these cases, the key challenge is to find optimal navigation strategies leading to a frequent and fast target finding with the least expense of resources.

In the past recent years, significant progress has been achieved in this direction analyzing intermittent search strategies~\cite{beni2011,beni2005}, optimal transport and related properties near surfaces~\cite{perez2019,schaar2015}, bacterial swimming in rugged chemoattractant gradients~\cite{Gosztolai2020}, optimal navigation in heterogeneous~\cite{yang2018,volp2017,yang2020,daddi2021} or scarce-resource~\cite{haeufle2016} environments, and in shear-flow problems~\cite{Liebchen2019}.
To date, however, little attention has been devoted to investigating optimal navigation strategies for the target-search problem in environments characterized by high energy barriers separating a target region from a reactant region.
If the particle needs to climb the barriers  by thermal fluctuations, the target search becomes a rare-event problem~\cite{Hanggi1990,wein2010,bolh2002,berryman2010}.
The self-propulsion of the agent can enhance the success rate significantly since the barriers can be overcome more easily~\cite{geis2016,roma2012,Woillez2019,militaru2021,Ebeling2005}, however, the particle may also explore different routes to find the target, thus changing  the typical search patterns~\cite{Zanovello2021}.

Here, we extend our previous analysis~\cite{Zanovello2021} for an active Brownian particle (ABP) in a double-well potential.
There, one-dimensional slices through parameter space have been discussed, while the goal of the current paper is to provide a full exploration of the model and to investigate if there exists an optimal navigation strategy for a given barrier height. 
Further objectives are to characterize the typical target-search patterns and to elucidate the interplay of motility parameters of the active agent and the underlying landscape for the target-search process.
A final aim of this work is to clarify whether these strategies are generic or dependent on the underlying landscape. 

First, we study the efficiency of an ABP to reach its target in a double-well potential, i.e. we analyze how the transition rates vary by changing the values of the activity and of the persistence of the particle.
Then we characterize the behavior of the ABP during the transitions in terms of other relevant quantities, such as the transition-path times (TPTs), which carry important information about the reactive dynamics~\cite{neup2016,chun2009,sega2007,zhan2007,cara2018,cara2020}, and the transition probability density, which also allows identifying the navigation patterns~\cite{wein2010,Metzner2006,Bartolucci2018}.
Finally, we explore the robustness of active-target-search features by extending this analysis to a different potential characterized by an additional metastable state.Active matter has received increasing attention in the past two decades~\cite{roma2012,Lauga_2009,bech2016} and has evolved to an interdisciplinary field, involving physics~\cite{roma2012,Fodor2016} biology~\cite{Cates_2012,Needleman2017,Lipowsky2005}, medicine~\cite{Henkes2020,Fodor2018}, and robotics ~\cite{marc2013,Erkoc2019}.
Examples of active matter systems include natural microswimmers, \textit{e.g.} bacteria, sperm cells, or amoebae~\cite{lauga2006,eise2006,wu2016}, but also artificially synthesized agents, such as the renowned Janus particles~\cite{howse2007} or micro- and nano-robots~\cite{bech2016}. 
Active particles have become a paradigmatic model to study the non-equilibrium dynamics of these systems, such as collective motion~\cite{vicsek2012,elge2015} or motility-induced phase separation~\cite{cates2015}, but are also of interest for the plethora of potential applications that they present already at the single-particle level.

Exploration and target search in complex environments are central problems in many realistic scenarios for such active agents.
For living organisms these tasks are a matter of survival, for example, in bacteria foraging nourishment~\cite{bech2016} or for sperm cells searching a fertile egg~\cite{eise2006}. 
Artificial swimmers have been envisaged to perform specific purposes on the microscale, such as targeted drug delivery~\cite{naah2013,patr2013} or water decontamination~\cite{Vilela2017}.
In all these cases, the key challenge is to find optimal navigation strategies leading to a frequent and fast target finding with the least expense of resources.

In the past recent years, significant progress has been achieved in this direction analyzing intermittent search strategies~\cite{beni2011,beni2005}, optimal transport and related properties near surfaces~\cite{perez2019,schaar2015}, bacterial swimming in rugged chemoattractant gradients~\cite{Gosztolai2020}, optimal navigation in heterogeneous~\cite{yang2018,volp2017,yang2020,daddi2021} or scarce-resource~\cite{haeufle2016} environments, and in shear-flow problems~\cite{Liebchen2019}.
To date, however, little attention has been devoted to investigating optimal navigation strategies for the target-search problem in environments characterized by high energy barriers separating a target region from a reactant region.
If the particle needs to climb the barriers  by thermal fluctuations, the target search becomes a rare-event problem~\cite{Hanggi1990,wein2010,bolh2002,berryman2010}.
The self-propulsion of the agent can enhance the success rate significantly since the barriers can be overcome more easily~\cite{geis2016,roma2012,Woillez2019,militaru2021,Ebeling2005}, however, the particle may also explore different routes to find the target, thus changing  the typical search patterns~\cite{Zanovello2021}.

Here, we extend our previous analysis~\cite{Zanovello2021} for an active Brownian particle (ABP) in a double-well potential.
There, one-dimensional slices through parameter space have been discussed, while the goal of the current paper is to provide a full exploration of the model and to investigate if there exists an optimal navigation strategy for a given barrier height. 
Further objectives are to characterize the typical target-search patterns and to elucidate the interplay of motility parameters of the active agent and the underlying landscape for the target-search process.
A final aim of this work is to clarify whether these strategies are generic or dependent on the underlying landscape. 

First, we study the efficiency of an ABP to reach its target in a double-well potential, i.e. we analyze how the transition rates vary by changing the values of the activity and of the persistence of the particle.
Then we characterize the behavior of the ABP during the transitions in terms of other relevant quantities, such as the transition-path times (TPTs), which carry important information about the reactive dynamics~\cite{neup2016,chun2009,sega2007,zhan2007,cara2018,cara2020}, and the transition probability density, which also allows identifying the navigation patterns~\cite{wein2010,Metzner2006,Bartolucci2018}.
Finally, we explore the robustness of active-target-search features by extending this analysis to a different potential characterized by an additional metastable state.

\section*{Methods}

\subsection*{Model and observables}
We study a microswimmer exploring a 2D energy landscape, starting from an initial region R (also called the reactant region) and trying to reach a target region T, separated by high energy barriers. The microswimmer is modeled as an active Brownian particle (ABP) described by Langevin equations
in It\^{o} discretization
\begin{subequations}
\begin{eqnarray}\label{eom}
\bm{r}_{i\!+\!1} &=& \bm{r}_{i} + v\, \bm{u}_{i} \, \Delta t - \mu \bm{\nabla} U(\bm{r}_{i}) \Delta t + \sqrt{2D\Delta t} \, \bm{\xi}_i\;,\\ \label{eom2}
\vartheta_{i\!+\!1} &=& \vartheta_{i} + \sqrt{2D_{\vartheta}\Delta t} \, \eta_i\;.
\end{eqnarray}
\end{subequations}
In the equations of motion (e.o.m\@) $\Delta t$ is the  time step and $\bm{r}_{i} = (x_{i},y_{i})$ denotes the position of the particle in the 2D plane at time $i \Delta t$.
The activity is introduced via the velocity term with constant modulus $v$ and a time-dependent orientation $\bm{u}_{i} = (\cos \vartheta_{i},\sin \vartheta_{i})$ evolving via a random walk of the angle $\vartheta_i$.
The mobility is denoted by $\mu$, $U(\bm{r})$ refers to the external potential, $D$ and $D_{\vartheta}$ are the translational and rotational diffusion coefficients.
Finally, the components of $\bm{\xi}_{i} = (\xi_{x,i},\xi_{y,i})$ and $\eta_{i}$ are independent centered Gaussian random variables of unit variance.

The productive trajectories for such a system, also called transition paths or reactive paths, are defined as those  trajectories leaving the R region and reaching the T region before they can re-enter in R.
Thus, the productive trajectories encode the information on how the transitions between the two states occur. 
In the case of high barriers the transitions become rare events. 
We characterize the reaction processes in terms of three observables.
First, we evaluate  the transition rate $r$, providing a measure on how frequently a particle undergoes a productive transition.
By considering extremely long trajectories including several reactive paths, the rate is computed as the inverse of the average time $t_{r}$ needed for a particle re-entering R (after having visited T) to reach T again.
The second quantity of interest is the average transition-path time $t_{\text{TPT}}$, i.e. the average time duration of the reactive paths.
Note that the transition path time ($t_{\text{TPT}}$) differs from the inverse rate ($t_{r}$) since the latter also accounts for the time spent in the reactant basin R and in non-productive paths, i.e. fragments of the whole trajectory leaving and re-entering R without meeting the target T. 
Last, we discuss the transition probability density $m(\bm{r})$ defined as the probability that a reactive trajectory visits a given position $\bm{r}$.

\subsection*{Parameters and simulation}
In the case of a passive Brownian particle the model contains only $\mu$ and $D$ as parameters, as well as the potential energy landscape $U(\bm{r})$ as input, while the orientational dynamics are uncoupled and can be ignored.
We reformulate the problem in terms of characteristic dimensionless parameters:
First, we note that even though our system is intrinsically out of equilibrium, it is still possible to express all energy scales in units of an effective thermal energy $k_B T := D/\mu$. Here, $T$ is physically interpreted as the temperature of the bath in the passive limit (i.e. for a standard overdamped Langevin dynamics).
Then we use the dimensionless potential $U(\bm{r})/k_B T$ to extract a characteristic length scale $L$ of the problem.
Here we choose $L := k_B T/F_{\text{max}}$ where $F_{\text{max}}$ is the maximum force along the minimum energy path linking the reactant region with the target region.
Thus $L$ can be interpreted as the length scale a passive particle can climb an energy landscape of slope $F_{\text{max}}$ by thermal fluctuations.
Finally, the time scale is set by $\tau := L^2/D$.

For passive particles the problem displays no further independent parameters and is uniquely specified, given the landscape $U(\bm{r})/k_B T$.
The self-propulsion adds the velocity $v$ and the rotational diffusion coefficient $D_{\vartheta}$ to the problem.
It is customary to introduce the P{\'e}clet number, $\text{Pe} := v L / D$, to quantify the importance of self-propulsion relative to the thermal fluctuations of the system.
The maximal force also translates  to  the maximal drift velocity $v_{\text{max}} = D F_{\text{max}}/k_B T$ the particle experiences on this minimum energy path.
The P{\'e}clet number can therefore be expressed as $\text{Pe} = v/v_{\text{max}}$, implying that an active particle with $\text{Pe} > 1$ can, in principle, take the minimum energy path without the help of thermal fluctuations.
However, orientational diffusion leads to a loss of persistence such that the active particle may change direction too rapidly to follow the minimum energy path.
The persistence length of an active particle $\ell := v/D_\vartheta$ is the typical distance on which the trajectories of a free active particle look straight.
We use the rotational diffusion coefficient to define  the second dimensionless parameter, $\ell^* := v/ D_\vartheta L = \ell/L$, which we refer to as persistence.
Given the shape of the dimensionless energy landscape $U(\bm{r})/k_B T$, the parameter space of the ABP is therefore two-dimensional and is spanned by $(\text{Pe}, \ell^*)$.
The case of a passive particle corresponds to the point $(\text{Pe}=0, \ell^* = 0)$ which will also serve as reference for the discussion.

\section*{Results and discussion}

In our previous work~\cite{Zanovello2021}, we have investigated a double-well potential and explored the parameter space 
for three sets of parameters: $(\text{Pe}=0,\ell^*=0)$ corresponding to the passive case, $(\text{Pe} \simeq 2,\ell^* \simeq 17)$, and $(\text{Pe} \simeq 4,\ell^* \simeq 34)$\footnote{Note that in our earlier work we use a different convention for the P{\'e}clet number, the current definition increases the former by a factor of $2L\sqrt{D_{\vartheta}/3D}$.}.
We showed that ABPs develop a counterintuitive target-search strategy which is markedly different from that adopted by a passive particle:
Instead of following the minimum energy path linking the reactant region with the target, ABPs exploit their directed self-propulsion to reach the target by surfing along high-energy regions of the potential landscape.
This surfing enhances the transition rates up to three orders of magnitude compared to the passive case, at the price of longer average TPTs. 
Furthermore, the dependence of these rates on the model parameters is non-trivial and displays a non-monotonic behavior at increasing activity.
Clearly, these observations depend on the interplay between the particle's activity $\text{Pe}$, its persistence $\ell^*$, and the shape of the landscape.

Here, we extend our previous study by considering the target-search problem for two paradigmatic energy landscapes and over a wide range of parameters $(\text{Pe}, \ell^*)$.
The goal is to identify the regions of parameter space where the target search becomes most efficient and characterize the features of the corresponding target-search strategy.
We study the system by direct integration of the equations of motion (\ref{eom}-\ref{eom2}).

\subsection*{Parameter space exploration for an ABP in a double-well}

\begin{figure}
\includegraphics[width=0.6\columnwidth]{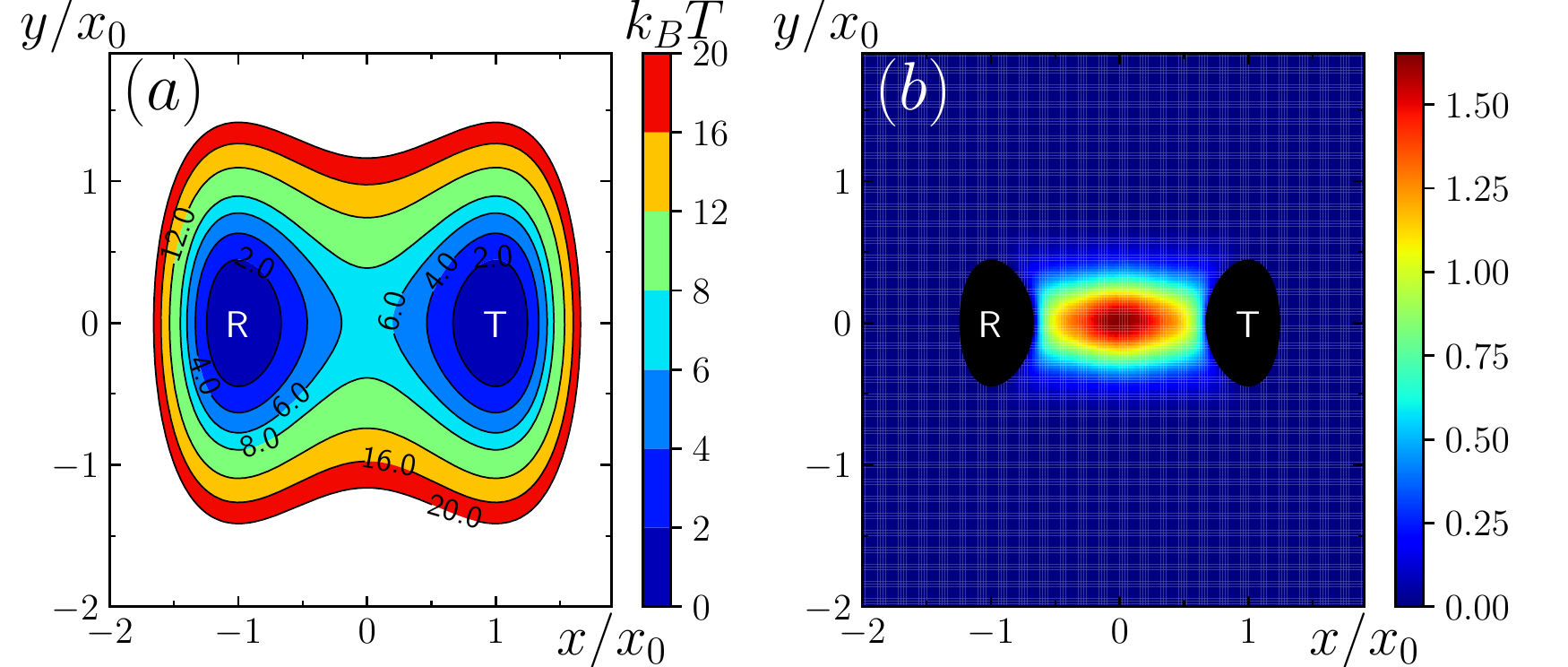}
\caption{(a) Representation of the double-well potential. The energy value in units of $k_{B}T$ is reported for each contour line. R identifies the reactant basin (defined as the region such that $U(x,y) \leq 2 k_{B}T$ and $x < 0$) and T represents the target region (defined by $U(x,y) \leq 2 k_{B}T$ and $x > 0$). (b) Transition probability density $m(\bm{r})$ for a passive particle in the double-well potential. The most visited configurations lie along the minimum energy path linking R and T, that crosses the barrier at the saddle point (0,0). In both panels $x_{0} \simeq 10 L$.}
\label{fig:dwpot}
\end{figure}

The simplest paradigmatic toy model used to study rare transition events is provided in terms of the double-well potential
\begin{equation}
U(\bm{r}) = k_{x} (x^{2}-x_{0}^{2})^{2} + \frac{k_{y}}{2}y^{2},
\label{eq:pot_dw}
\end{equation}
with $\bm{r} = (x,y)$.
Despite its simplicity, this landscape has already proven to offer a rich transition phenomenology~\cite{Zanovello2021}.
The potential displays two minima, located at $(\pm x_0, 0)$, and the minimum energy path is along the $x$-direction with a saddle point at $(0,0)$, see Fig.~\ref{fig:dwpot}(a). 
The parameter $k_x x_0^4$ then corresponds to the barrier height, the ratio $k_y/k_x x_0^2$ characterizes the shape of the potential, while $x_0$ is a geometric length of the landscape.
The maximal force along the minimum energy path occurs at $(\pm x_0/\sqrt{3},0)$ and evaluates to $F_\text{max} = 8 k_x x_0^3/3\sqrt{3}$. 
To fix the potential once and for all, we choose the barrier as $k_x x_0^4 = 6.5  k_B T$ which makes the transition a rare-event for passive particles.
It follows that $F_\text{max} \simeq 10 k_BT/x_0$.
To fix the shape of the potential energy landscape we use $k_y x_0^2 =20 k_B T $, which means that a passive particle displays typical excursions in the $y$-direction of order $\sqrt{k_B T/ k_y} = x_0/\sqrt{20}$. 
The characteristic length scale that we use to define the P{\'e}clet number and the persistence then reads $L =  3\sqrt{3}k_{B}T/8k_{x}x_{0}^{3}$.
We define the R region by $U(x,y) \leq 2 k_{B}T$ and $x < 0$ and the T region by $U(x,y) \leq 2 k_{B}T$ and $x > 0$.

\begin{figure*}
\centering
\includegraphics[width=1.\columnwidth]{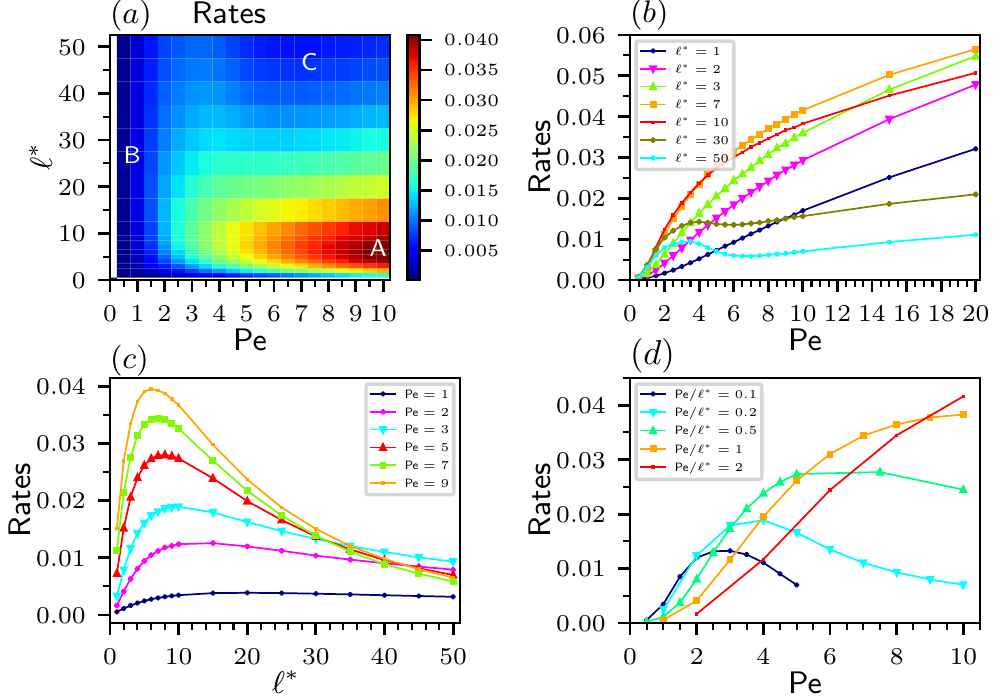}
\caption{(a) Transition rates in the double-well as a function of $\text{Pe}$ and $\ell^*$. The most favorable parameters for the transition rates are identified in region A, while two distinct unfavorable regions, B and C, show much smaller transition rates. 
(b) Transition rates represented as a function of $\text{Pe}$ for different values of $\ell^{*}$. 
In the investigated parameter space, the highest rates are observed for $ 3 \lesssim \ell^{*} \lesssim 10$.
(c) Transition rates represented as a function of $\ell^{*}$ for different values of $\text{Pe}$. 
For each $\text{Pe}$, a value of $\ell^{*}$ that maximizes the rates is found. The larger $\text{Pe}$ becomes, the smaller this $\ell^{*}$  will be. 
(d) Transition rates represented as a function of $\text{Pe}$ for different values of the ratio $\text{Pe}/\ell^{*}$.
The trend shows that for each ratio $\text{Pe}/\ell^{*}$ a value of $\text{Pe}$ that maximizes the rates exists. This value becomes larger with increasing $\text{Pe}/\ell^{*}$.}
\label{fig:ratesdw}
\end{figure*}

\begin{figure}[h]
\centering
\includegraphics[width=0.6\columnwidth]{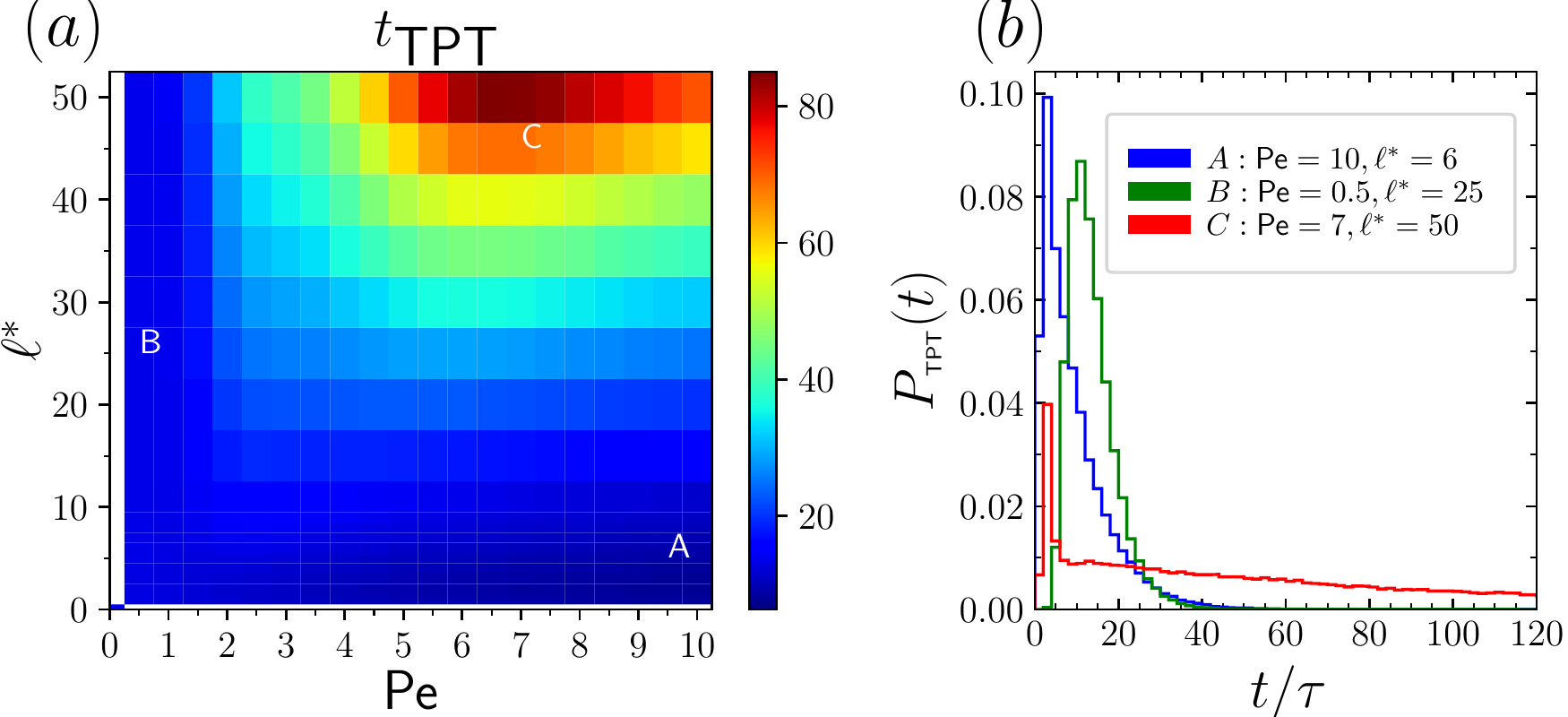}
\caption{(a) Average transition path times $t_{\text{TPT}}$ as a function of $\text{Pe}$ and $\ell^{*}$ in the double-well potential. Both regions A and B are characterized by short average TPTs, while region C displays much longer times. (b) Distribution of transition-path times (TPTs) at different P{\'e}clet numbers and persistence. Each distribution is obtained from $10^5$ different TPTs.}
\label{fig:tptdw}
\end{figure}

\begin{figure*}
\centering
\includegraphics[width=1.\columnwidth]{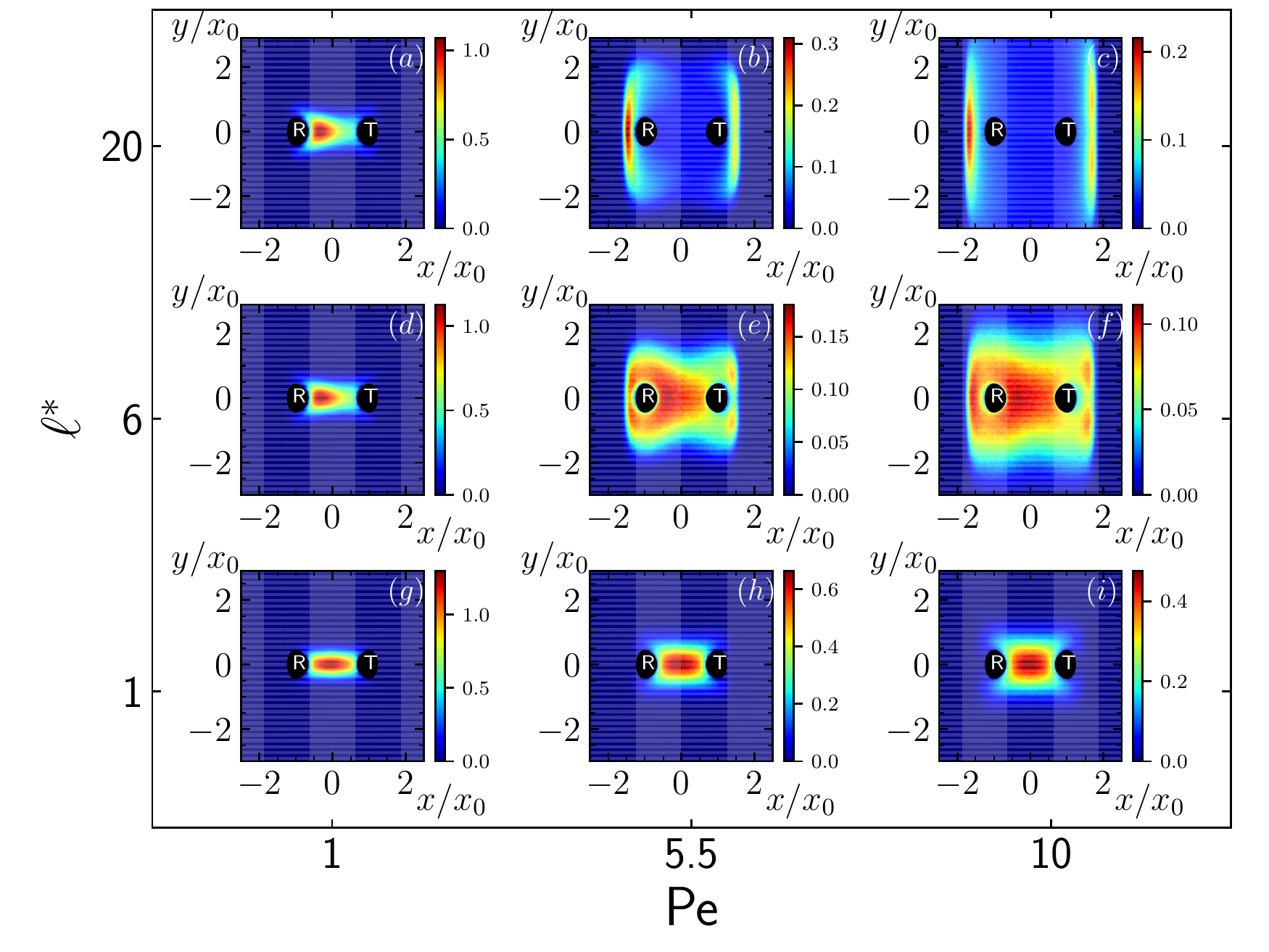}
\caption{Transition probability density $m(\bm{r})$ for different values of $\text{Pe}$ and $\ell^{*}$ in the double-well potential. 
At small values of $\text{Pe}$ (a,d,g) the transition probability density resembles the one of a passive particle, with most frequently visited configurations located along the minimum energy path. 
This similarity is smeared out with an increase in $\ell^{*}$.
An increase in $\text{Pe}$ leads to the exploration of larger portions of the transition region. 
At small values of $\ell^{*}$ (h,i) the distributions still resemble the one of a passive particle but with larger diffusion coefficient the more $\text{Pe}$ increases. 
At intermediate values of $\ell^{*}$ (e,f) the active particle is able to efficiently explore the transition region, while an increase in $\ell^{*}$ (b,c) causes the particle to take long detours at the boundaries.}
\label{fig:RPdw}
\end{figure*}

Passive particles, $(\text{Pe}=0,\ell^*=0)$, display a transition rate of about $8\cdot10^{-5} \tau^{-1}$ and always cross the barrier in the proximity of the minimum energy path, see Fig.~\ref{fig:dwpot}(b).
Active particles instead display higher rates and a rich transition phenomenology.
The most favorable transition rates are found at intermediate values of $\ell^{*}$ and large values of $\text{Pe}$ (region A in Fig.~\ref{fig:ratesdw}(a)), characterized by values of about $4\cdot10^{-2} \tau^{-1}$.
In this region, the high activity ($\text{Pe} \gtrsim 5$) allows the particle to easily overcome the energy barrier and, at the same time, the intermediate persistence ($2 \lesssim \ell^* \lesssim x_0/L \simeq 10$) ensures that the particle properly explores and efficiently navigates in the double-well, without surfing too long on the high-energy regions.
These observations are confirmed by a systematic statistical analysis of the TPTs, see Fig.~\ref{fig:tptdw}, and of the transition probability density, see~\ref{fig:RPdw}(e-f).
Outside of region A, two unfavorable regions emerge (region B at small values of $\text{Pe}$ and region C at large $\text{Pe}$ and large $\ell^{*}$).
Region B and C show low rates for different reasons:
Region B is characterized by a small self-propulsion which implies that the particle will not be able to efficiently cross the barrier, leading to low rates.
In a similar fashion to a passive particle, ABPs belonging to region B also display short TPTs, see Fig.~\ref{fig:tptdw}, and a transition probability density mainly spread around the minimum energy path, see Fig.~\ref{fig:RPdw}(a,d,g) and Fig.~\ref{fig:dwpot}(b) as reference for the passive behavior.
Interestingly, when comparing regions A and B, the average TPT decreases with the activity of the particle as also observed in one-dimensional systems~\cite{carl2018}.
On the other hand, particles in region C can easily cross the barrier, thanks to their large self-propulsion, but the persistence of the propulsion direction forces them to take extremely long-lasting transitions along the confining boundaries of the potential, far from the target basin, see Fig.~\ref{fig:RPdw}(b-c).
Concomitantly, the TPT distribution shows a long-time tail and the average TPTs in this region are much larger than in the rest of the parameter space, see Fig.~\ref{fig:tptdw}, and represent the largest contribution to $t_{r}$.
However, in region C, the $t_{\text{TPT}}$ distribution still displays a peak at short times, corresponding to particles that, when crossing the barrier, already have a favorable direction pointing towards the target.
Region C is separated from region B by a section of the parameter space with more favorable rates, as also revealed by considering linear slices of the parameter space at fixed and large persistence $\ell^{*}$, see Fig.~\ref{fig:ratesdw}(b).
This intermediate region has no counterpart in the behavior of the average TPT (Fig.~\ref{fig:tptdw}(a)) and represents those particles with a self-propulsion that is sufficient to effectively climb over the barrier but is not so large to make them, in combination with the high persistence, spend long fractions of their trajectories trying to climb the confining potential walls.

Fig.~\ref{fig:ratesdw} suggests criteria to optimize the target-search strategy also in the presence of constraints.
In particular, at fixed $\ell^{*}$, the best action to enhance the transition rates would be to increase $\text{Pe}$.
In fact, except for the already discussed initial non-monotonic behavior of the rates observed at large $\ell^*$ values, the transition rates grow with increasing activity, see Fig.~\ref{fig:ratesdw}(b).
This observation may appear counterintuitive at first:
While at very low $\text{Pe}$ an increase of the activity obviously leads to an increase of the rates because climbing the barrier becomes easier, at large $\text{Pe}$ we would naively expect that the higher the activity is, the longer is the time spent in the high-energy regions.
However, a more careful inspection of the TPTs behavior (Fig.~\ref{fig:tptdw}(a)) shows that, at fixed $\ell^*$, after an initial growth they start to decrease with increasing P{\'e}clet number and this also affects the transition rates.
The reason for this trend is that, at fixed $\ell^*$, $\text{Pe}$ is directly proportional to the rotational diffusion coefficient $D_{\vartheta}$.
Thus, enhancing the activity also enhances the rotational diffusion with a consequent decrease of the time spent surfing the energy walls.

If instead a particle is unable to enhance its self-propulsion, it can still increase its odds of a successful target search by tuning its persistence length in response to the features of the environment.
In fact, from Fig.~\ref{fig:ratesdw}(c) one infers that for a fixed value of $\text{Pe}$ there will be a value of $\ell^{*}$ that maximizes the rate, this value becoming smaller when $\text{Pe}$ increases.
The reason for this tendency is that if the activity is too small to cross quickly the energy barrier, the particle will benefit more from a larger value of $\ell^{*}$ to keep its swimming orientation longer, so that all the swimming efforts are summed in the same direction for larger times.

Lastly, if the particle self-propulsion is adjustable but its rotational diffusion coefficient is not, i.e. the constraint consists in having a constant value of the ratio $Pe/\ell^{*}$, the rates can be maximized by choosing a specific value of the self-propulsion for that given ratio (see Fig.~\ref{fig:ratesdw}(d)).
In particular, the best rates are found at larger values of this ratio, and the corresponding optimal $\text{Pe}$ increases with the value of the ratio.

\subsection*{Parameter space exploration for an ABP in the Brown-M\"uller potential}

\begin{figure}[h]
\centering
\includegraphics[width=0.6\columnwidth]{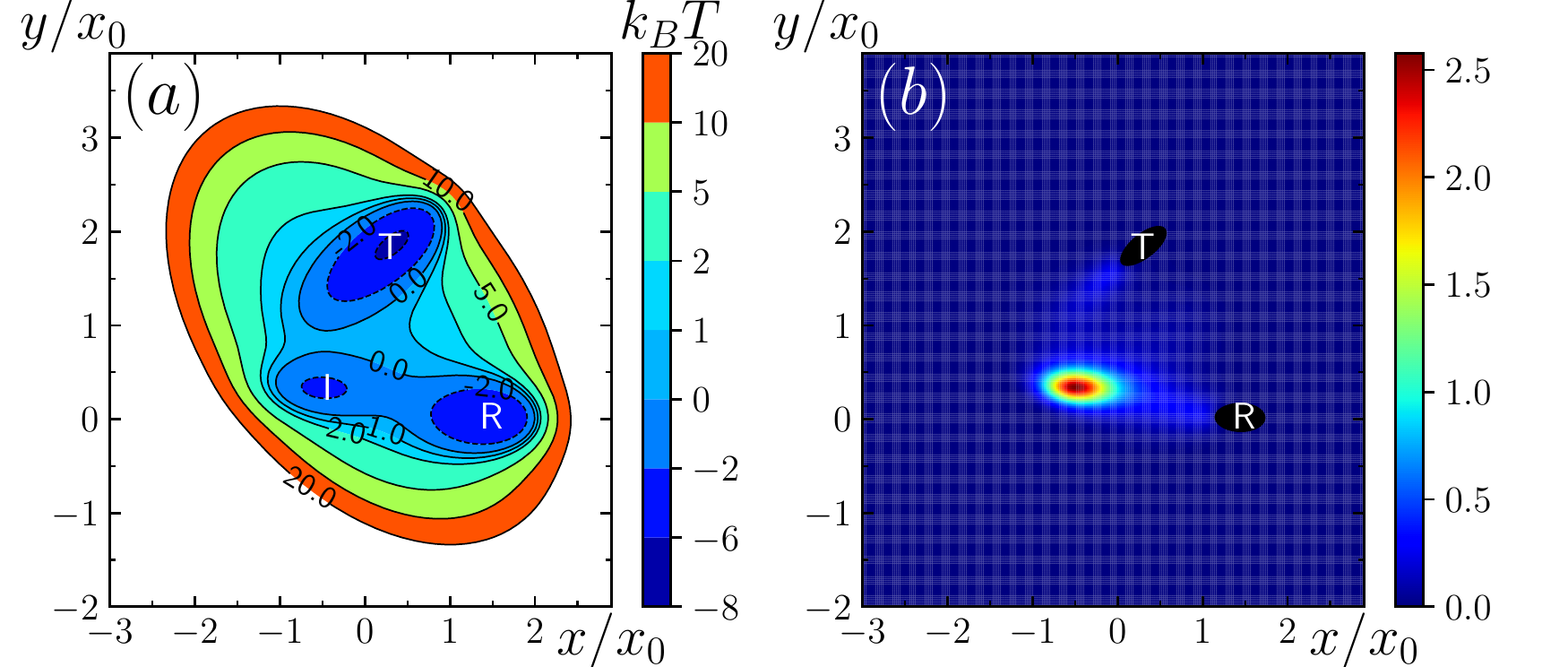}
\caption{(a) Representation of the Brown-M\"uller potential. 
The energy value in units of $k_{B}T$ is reported for each contour line. 
R identifies the reactant basin (defined as the region such that $U(x,y) \leq -4.5 k_{B}T$ and $y < 0.7$), 
T represents the target region (defined by $U(x,y) \leq -5.5 k_{B}T$ and $y > 0.7$), and I is an intermediate metastable state. (b) Transition probability density $m(\bm{r})$ for a passive particle in the Brown-M\"uller potential. The most visited configurations lie along the minimum energy path linking R and T, that passes across the intermediate state I where the particle spends most of its transition time. In both panels $x_{0} = 10 L$.}
\label{fig:bmpot}
\end{figure}

We define the R region by $U(x,y) \leq -4.5 k_{B}T$ and $y < 0.7$ and the T region by $U(x,y) \leq -5.5 k_{B}T$ and $y > 0.7$.

\begin{figure*}
\centering
\includegraphics[width=1.\columnwidth]{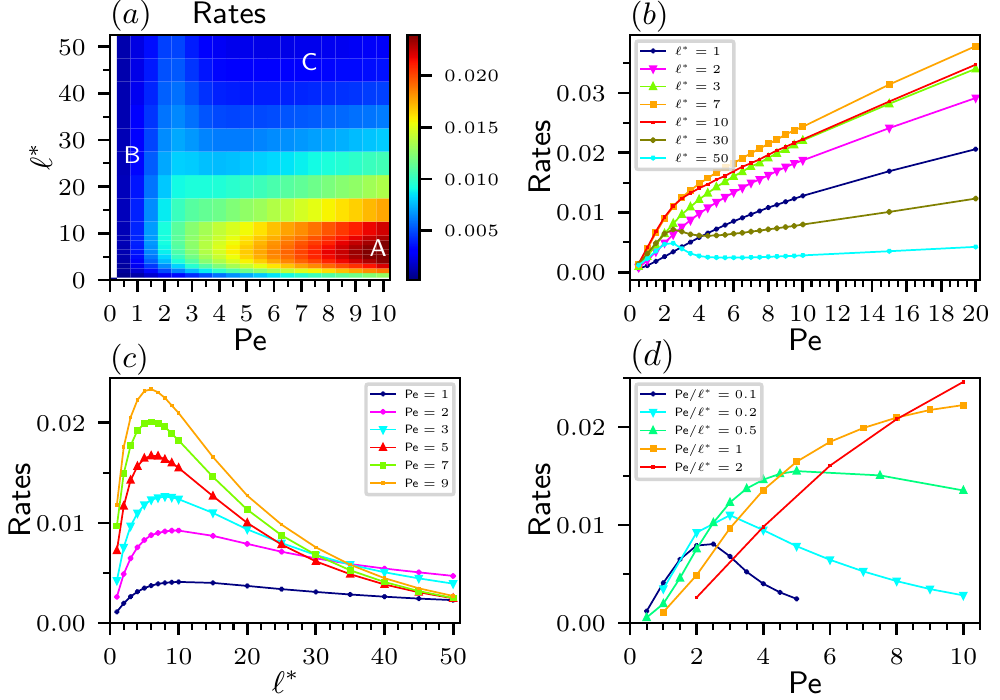}
\caption{(a) Transition rates in the Brown-M\"uller potential as a function of $\text{Pe}$ and $\ell^*$. The most favorable parameters for the transition rates are identified in region A, while two distinct unfavorable regions, B and C, show much smaller transition rates. 
(b) Transition rates represented as a function of $\text{Pe}$ for different values of $\ell^{*}$.
In the investigated parameter space, the highest rates are observed for $3 \lesssim \ell^{*} \lesssim 10$.
For values of $\ell^{*}$ between $3$ and $10$ the highest rates are observed, while for smaller or larger values of $\ell^{*}$ the rates are lower. Again, after an initial part of the curves, the transition rates increase linearly with $\text{Pe}$. 
(c) Transition rates represented as a function of $\ell^{*}$ for different values of $\text{Pe}$. For each $\text{Pe}$, a value of $\ell^{*}$ that maximizes the rates is found. The larger $\text{Pe}$ becomes, the smaller this $\ell^{*}$ will be. 
(d) Transition rates represented as a function of $\text{Pe}$ for different values of the ratio $Pe/\ell^{*}$.}
\label{fig:ratesmuller}
\end{figure*}

An important question to address is whether some of these observations are strongly influenced by the choice of a simple double-well potential, that is symmetric along both the dimensions and has a single barrier separating the R and T basins.
To address this issue, we carry out a similar analysis for a different energy landscape, namely the Brown-M\"uller potential~\cite{Muller1979}:
\begin{equation}
    U(x,y) = \sum_{i=1}^{4} K_{i} e^{[a_{i}(x-x_i)^{2}+b_{i}(x-x_i)(y-y_i)+c_{i}(y-y_i)^{2}]} \; ,
\end{equation}
where we set $K_i/k_BT = (-9.69,-4.41,-8.81,0.88)$, $x_i/x_0 = (1.7,-0.7,0.5,0)$, $y_i/x_0 = (0,0.3,2,1)$, $a_{i}/x_0^2 = (-1.2,-1.2,-3,0.7)$, $b_{i}/x_0^2 = (0,0,6,0.6)$, and $c_{i}/x_0^2 = (-6,-8.5,-5,0.7)$.
We define the R region by $U(x,y) \leq -4.5 k_{B}T$ and $y < 0.7$ and the T region by $U(x,y) \leq -5.5 k_{B}T$ and $y > 0.7$.
The energy barrier along the minimum energy path linking R and T exhibits a maximum height of about $6 k_{B}T$ measured from the bottom of the R basin, while the R and T basins have a depth of about $1.3 k_{B}T$ and $1.1 k_{B}T$ respectively.
By construction, $F_{\text{max}} \simeq 10 k_BT/x_0$ so that the characteristic length of the system $L$ is the same as in the double-well.
This energy landscape is markedly asymmetric (Fig.~\ref{fig:bmpot}), there is a third intermediate metastable state I, and its minimum energy path linking R to T is curved and meets two barriers.

Notwithstanding the differences between the two considered potentials, the dependence of transition rates on the parameters remains quite similar, compare Fig.s~\ref{fig:ratesdw}(a) and~\ref{fig:ratesmuller}(a).
The passive particle, $(\text{Pe}=0,\ell^*=0)$, sets a reference rate of $2.4\cdot10^{-4}\tau^{-1}$, which is about three times larger than in the double-well, compatibly with the fact that a smaller barrier height separates the R and T basin in the Brown-M\"uller potential.
In the Brown-M\"uller potential the most favorable rates are about $2\cdot10^{-2}\tau^{-1}$, about half of those observed in the double-well.
Given the smaller barrier height of this landscape with respect to the double-well, at first glance this decrease in the rates may seem counterintuitive.
However, active particles with sufficient self-propulsion are not significantly affected by the height of the barrier and, in this case, the rates decrease due to the reduced area of the target region, see Fig.s~\ref{fig:dwpot}(b) and~\ref{fig:bmpot}(b) for comparison.

\begin{figure}[h]
\centering
\includegraphics[width=0.6\columnwidth]{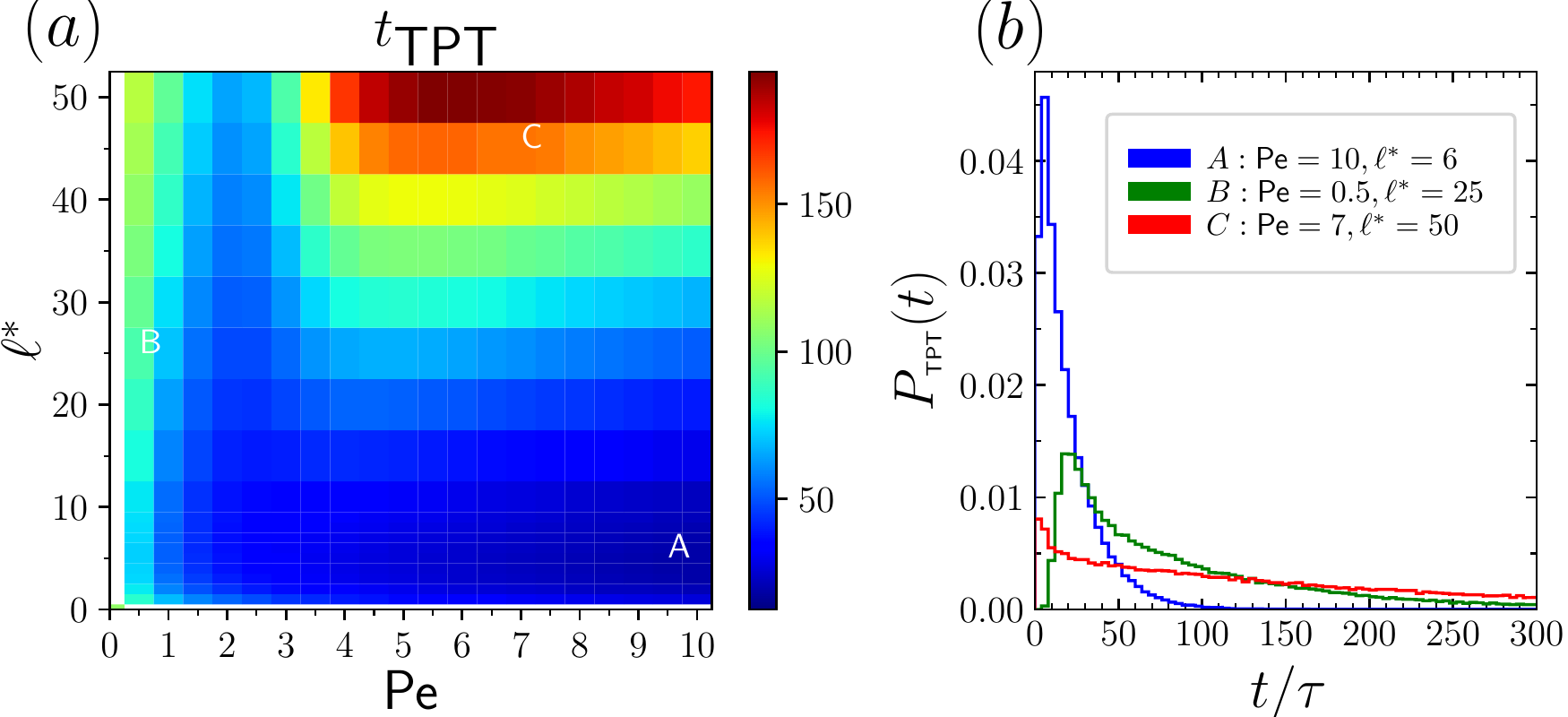}
\caption{(a) Average transition path times $t_{\text{TPT}}$ as a function of $\text{Pe}$ and $\ell^{*}$ in the Brown-M\"uller potential. Region A is characterized by short average TPTs, while regions B and C display much longer times. (b) Distribution of transition-path times (TPTs) at different P{\'e}clet numbers and persistence. Each distribution is obtained from $10^5$ different TPTs.}
\label{fig:tptmuller}
\end{figure}

\begin{figure*}
\centering
\includegraphics[width=1.\columnwidth]{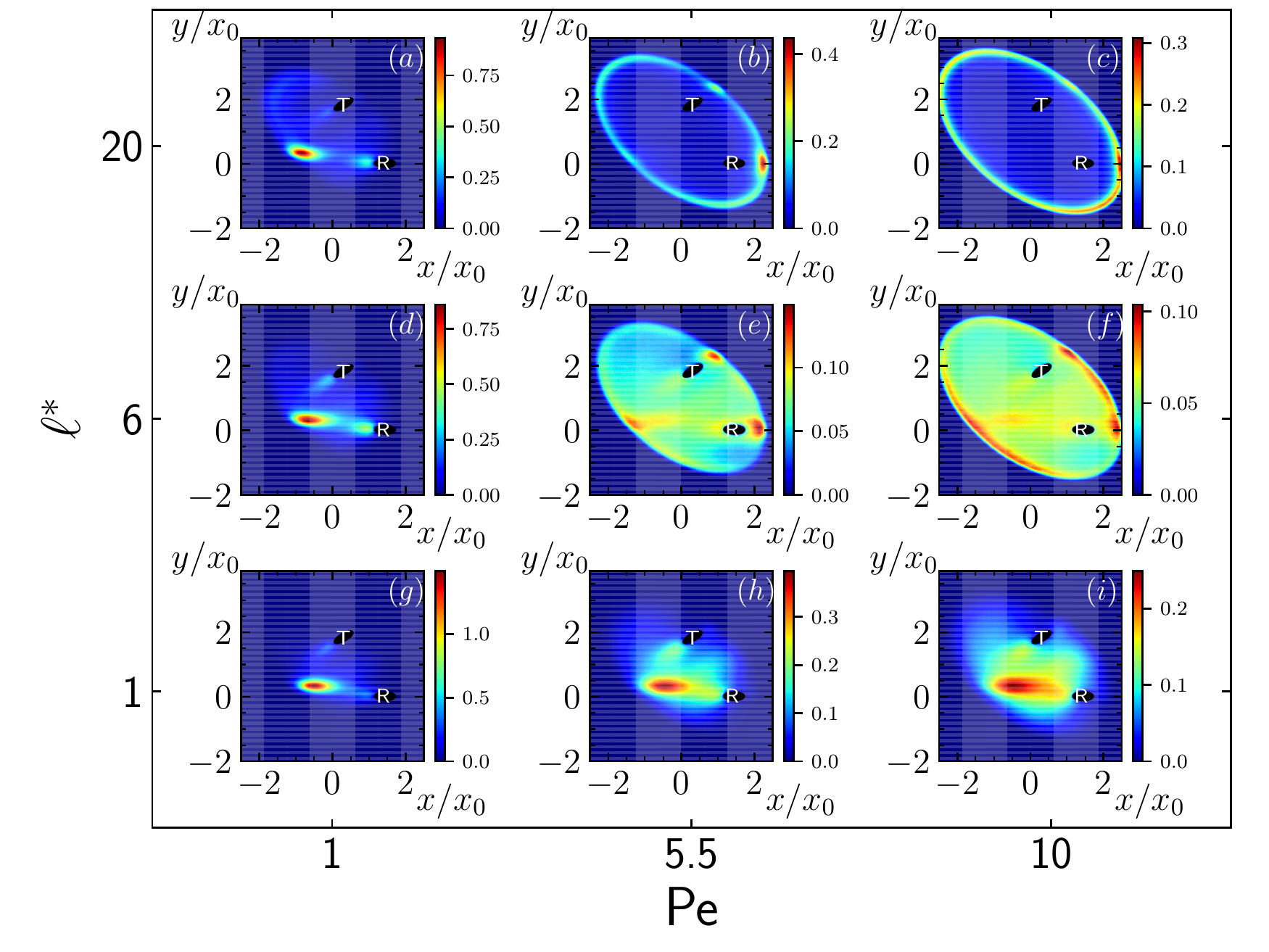}
\caption{Transition probability density $m(\bm{r})$ for different values of $\text{Pe}$ and $\ell^{*}$ in the Brown-M\"uller potential. 
At small values of $\text{Pe}$ (a,d,g) the transition probability density resembles the one of a passive particle, with most frequently visited configurations located along the minimum energy path and long times spent in the metastable state I.
This similarity diminishes with an increase in $\ell^{*}$. 
An increase in $\text{Pe}$ leads to the exploration of larger portions of the transition region. 
At small values of $\ell^{*}$ (h,i) the distributions still resemble those of a passive particle but with larger diffusion coefficient the more $\text{Pe}$ increases . 
At intermediate values of $\ell^{*}$ the active particle is able to efficiently explore the transition region (e,f), while a further increase in $\ell^{*}$ (b,c) causes the particle to take long detours at the boundaries.}
\label{fig:RPmuller}
\end{figure*}

Again, the transition rates display a favorable region (region A in Fig.~\ref{fig:ratesmuller}(a)) at intermediate values of $\ell^{*}$ and large values of $\text{Pe}$, while the two unfavorable regions B and C are found at small $\text{Pe}$ and at large $\text{Pe}$ and $\ell^{*}$ respectively.
Also similarly to the double-well case, when $\ell^{*}$ becomes too small, i.e. when the particle changes too quickly its orientation, the rates decrease because the particle behaves similarly to the passive particle (even if with a larger effective diffusion coefficient) and spends a consistent fraction of the transition time in the intermediate state I (see Fig.~\ref{fig:RPmuller}(g-i) and Fig.~\ref{fig:bmpot}(b) as reference for the passive behavior).
Finally, regions B and C in Fig.~\ref{fig:ratesmuller}(a) show lower transition rates for the same reasons that cause the emergence of this behavior in the double-well.
The particles in region B have a small value of the self-propulsion, thus similarly to the agents in region B of the double-well they are not able to easily climb over the barriers.
In this case, however, they spend a large fraction of their transition time in the metastable state I, that is not present in the double-well.
This is reflected in longer average TPTs than in region A, see Fig.~\ref{fig:tptmuller}(a), in striking contrast to the double-well, and in the transition probability densities in Fig.~\ref{fig:RPmuller}(a,d,g), that highlight the permanence in the metastable state.
The emergence of a long-time tail in the TPT distribution in region B, see Fig.~\ref{fig:tptmuller}(b), together with the difficulty in initiating a successful transition by climbing the barriers, generates the low rates observed in the region.
On the other hand, particles belonging to region C are characterized by a large activity, that allows them to easily climb the barriers, and a large persistence, that makes them surf along the boundaries of the system for long times before being able to change orientation and reach the T state (Fig.~\ref{fig:RPmuller}(b,c)).
This generates long TPTs (see Fig.~\ref{fig:tptmuller}), that decrease at very large $\text{Pe}$, in a similar fashion to the double-well, due to the fact that increasing $\text{Pe}$ at fixed $\ell^*$, automatically means to increase also the rotational diffusion coefficient $D_{\vartheta}$.
In contrast to the double-well, however, regions B and C are well separated not only in the rates, by the presence of a local maximum, but also in the average TPTs, where a local minimum separates the region in which the long TPTs are due to a long permanence in the intermediate state (region B) from the region in which long TPT are due to long detours at the outskirts of the system.

All these observations suggest the general rule that active particles achieve a more effective navigation when their activity is large and the persistence length is such that it contributes to overcome the energy barrier but it is not too large to become a counter-productive factor.
They also suggest that additional metastable states do not affect this rule, at least as long as their depth does not exceed that of the reactant and the target basin.

In complete analogy to the case of the double-well, our results show how to optimize the target-search success also in the presence of constraints on the two control parameters.
If the particle is unable to modify its persistence, the better chance to enhance the transition rates is by increasing the self-propulsion, even if, at very large persistence and at small activity, this may initially lead to the opposite result, due to the non-monotonicity of rates with increasing $\text{Pe}$, see Fig~\ref{fig:ratesmuller}(b).
At fixed self-propulsion, one can still optimize the transition rate by picking the optimal $\ell^{*}$ value, as can be seen in Fig.~\ref{fig:ratesmuller}(c).
Again, this optimal $\ell^{*}$ will become smaller the larger is the activity of the ABP.
An equivalent behavior to the double-well is observed also in the case of a particle constrained to a constant value of the ratio $Pe/\ell^{*}$, for which the rates can be maximized by picking the optimal $\text{Pe}$ once the ratio is fixed.
Once more, the best rates are then achieved when the ratio is as large as possible, and the $\text{Pe}$ value that optimizes the rates increases with the value of the ratio, see Fig.~\ref{fig:ratesmuller}(d).

\section*{Conclusions}

We have studied the behavior of an ABP searching for a target in an energy landscape, aiming at finding the region of the parameter space where the target search becomes most efficient and at describing the corresponding target-search strategy.
To characterize the transition process we have analyzed, as a function of the ABP's self-propulsion and persistence, a set of observables: The transition rate, the transition path time (TPT) and the transition probability density.
We studied the problem both in a simple double-well potential, as well as in a more complex landscape, the Brown-M\"uller potential, that presents evident asymmetries and an intermediate metastable state.
We found that the target-finding rate can be optimized by tuning the self-propulsion and the persistence in response to external stimuli, which in our model are set by the energy landscape that the ABP is navigating in.
The most favorable region in parameter space is nearly independent of the details of the energy landscape and the corresponding strategy adopted by the agent is based on an efficient exploration of the transition region.
This is guaranteed by a self-propulsion sufficiently large to easily overcome the energy barriers and by an intermediate level of the persistence, thus avoiding passive-like trajectory shapes (small $\ell^{*}$) or too long detours in high-energy regions at the outskirts of the system (large $\ell^{*}$).
Even though outside the most favorable region the rates are lower, active particles always display better performances in reaching the target with respect to passive ones.
The similar qualitative behavior of the rates between the two potential landscapes suggests the generality of these findings, and, additionally, that the most efficient exploration strategy for an ABP navigating rugged energy landscapes can be achieved following a single protocol when tuning the parameters.
Finally we showed that if the active agent is subject to some constraints, e.g. it cannot modify its persistence, its self-propulsion or its rotational diffusion coefficient, the rates can still be optimized by controlling the remaining control parameter.

In contrast to the rates, the TPTs are more sensitive to the details of the energy landscape.
While they show similar features in the two landscapes for large values of the activity, significant differences emerge when the self-propulsion is low.
In fact, in this regime the presence of the additional metastable state in the Brown-M\"uller potential leads to much longer TPTs than in the double-well, compatibly with the fact that in these conditions ABPs behave similarly to passive particles, thus spending a large fraction of their transition time in the intermediate state.

From the characterization of the transition probability density emerges one of the peculiar features of the ABPs: consistently with previous studies of the steady-state probability distributions for ABPs in confining environments~\cite{Dauchot2019,Pototsky2012,Takatori2016,malakar2020}, when activity and persistence are sufficiently large, they tend to spend long fractions of their trajectories at the boundaries of the system.
Consequently, they experience difficulties in reaching a specific region far from the boundaries similarly to what happens in the case of an ABP looking for a target located far from the rigid boundaries of a confining environment~\cite{wang2016}.

To summarize our findings, we have shown that the best strategy adopted by ABPs to find a target involves short surfs along high-energy regions.
The addition of an intermediate metastable state in the system has allowed us to gather some insight on how ABPs explore rugged environments and reach a target therein.
In particular, while the optimal target search patterns reflect the presence of additional minima, the rates and the TPTs are virtually unaffected by these kinetic traps.
Hence, activity proves to be once again an advantageous feature in the exploration of complex energy landscapes and in target-search problems.

Possible applications and further works may aim at synthesizing active agents able to regulate their activity and persistence in response to the external stimuli to achieve the most efficient navigation in rugged energy landscapes.

\begin{acknowledgments}
This work was supported by the Austrian Science Fund (FWF): Grant No. P28687-N27.
\end{acknowledgments}


%

\end{document}